# Droplet deformation and fragmentation by ultra-short laser pulses

M. S. Krivokorytov[1,2], A. Yu. Vinokhodov[1,2], Yu. V. Sidelnikov[1,3], V. M. Krivtsun[1,3], V.V. Medvedev[1,3], V.O.Kompanets[3], A.A. Lash[1,2] and K.N. Koshelev[2,3]

1 – RnD-ISAN, Promyshlennaya Str., 1A, 142191, Troitsk, Moscow, Russia; 2 – EUV Labs, Sirenevy bulvar Str., 1, 142191, Troitsk, Moscow, Russia; 3 – Institute for Spectroscopy RAS, Fizicheskaya Str., 5, 142090, Troitsk, Moscow, Russia.

**Abstract**

We report on the experimental studies of the deformation and fragmentation of liquid metal droplets by picosecond and subpicosecond laser pulses. The experiments were performed with laser irradiance varying in $10^{13}$-$10^{15}$ W/cm$^2$ range. The observed evolution of the droplet shape upon the impact dramatically differs from the previously reported for nanosecond laser pulses. Instead of flattening the droplet undergoes rapid asymmetric expansion and transforms into a complex shape which can be interpreted as two conjunct spheroid shells and finally fragments. We explain the described hydrodynamic response to the ultra-short impact as a result of the propagation of the laser-induced convergent shockwave through the volume of droplet.

**Manuscript**

The interaction of laser pulses with matter generates a rich variety of hydrodynamic phenomena that are of great fundamental interest and simultaneously form a basis for multiple applications. Examples of such phenomena are ejection of liquid droplets and jets from laser irradiated thin films [1-4], laser-induced cavitation in liquids confined in capillary channels [5-7], and deformation and fragmentation of sessile and mobile liquid droplets induced by laser pulses [8-11]. The latter phenomena attracted an increasing attention of researchers in the recent years [9-11]. The current demand for understanding of the liquid drop deformation by laser pulse impact is very high due its direct relevance for the application in laser-produced plasma sources of extreme ultraviolet radiation for the next-generation photolithography [12-14]. These sources make use of liquid tin (Sn) droplets acting as fuel-delivering targets for laser-produced plasma. The droplets are irradiated by double laser pulses. The first pulse of moderate intensity is used to optimize the target shape before shooting it with the second pulse of high intensity which is used to create the plasma emitting EUV photons. Optimization of the target shape requires the detailed understanding of the hydrodynamic response of droplets to the impact of laser pulse.

Previous studies addressed the phenomena that occur when irradiating liquid droplets with laser pulses of tens nanosecond duration. The studies were performed for dyed water drops in the regime of laser evaporation of the surface [9] and for liquid tin drops at higher irradiance in the regime of laser ablation of the surface [11]. Klein et al reported that the dynamics of droplet response shot by nanosecond laser pulses is similar to that observed when liquid droplet endure mechanical impact on solid substrate [9]. To be more precise, under such impact conditions droplets flatten and expand in the direction perpendicular to the laser beam axis before it either retracts at low laser energies or fragments at high energies. In this paper we report on the droplet shaping by laser pulses with duration ($\tau$) ranging from 50 fs to 5.3 ps. The observed hydrodynamic response of drops to such ultra-short laser impact dramatically differs from the previously reported for nanosecond laser pulses. As a result of the impact, the droplet is transformed into a complex-shaped hollow structure which undergoes asymmetrical expansion and eventually fragments. We interpret the observed scenario of the droplet deformation as the result of propagation of the intense laser-induced shockwave through the body of droplet.

**In the experiments we used droplets of liquid Sn-In eutectic alloy with 48% - 52% mass stoichiometry.** Melting temperature of the alloy is 119°C, which is almost two times lower than the melting temperature of pure tin and indium. The experiments were performed in vacuum chamber with pressure of residual gases < $10^{-4}$ mbar. Working temperature was 140°C. Low temperature allows easier experimental operation, but all other physical properties of alloy (density, surface tension etc.) are similar to pure tin. Droplets were generated using a commercial droplet dispenser MJ-SF-01 by MicroFab Technologies. The diameter of the droplets was 60 um. The velocity of droplets was 3 m/s. The droplets were shot by pulses generated by Ti:Sapphire laser system. The laser system consisted of Ti:Sapphire generator (Spectra Physics Tsunami, λ = 780 – 820 нм), regenerative amplifier (Spectra Physics SpitFire) and corresponding pumping lasers (Spectra Physics Millennia eV и Spectra Physics Empower). The regenerative amplifier allowed us changing pulse duration (τ) from 50 fs and up to 5.3 ps. The pulse duration was measured by means of autocorrelation function with PulseScout autocorrelometer. The experiments were performed at the maximum achievable pulse energy (E) of 2.3 mJ. For the visualization of droplet shape evolution upon the laser impact we used the stroboscopic shadowgraphy technique employing the pulsed laser back-lighting of the object. For that we used diode laser IL30C which generates 30 ns pulses at the wavelength of 850 nm. The shadow images were recorded by using long distance microscope and a CCD camera. The details on experimental setup, synchronization and diagnostics are described in [19].

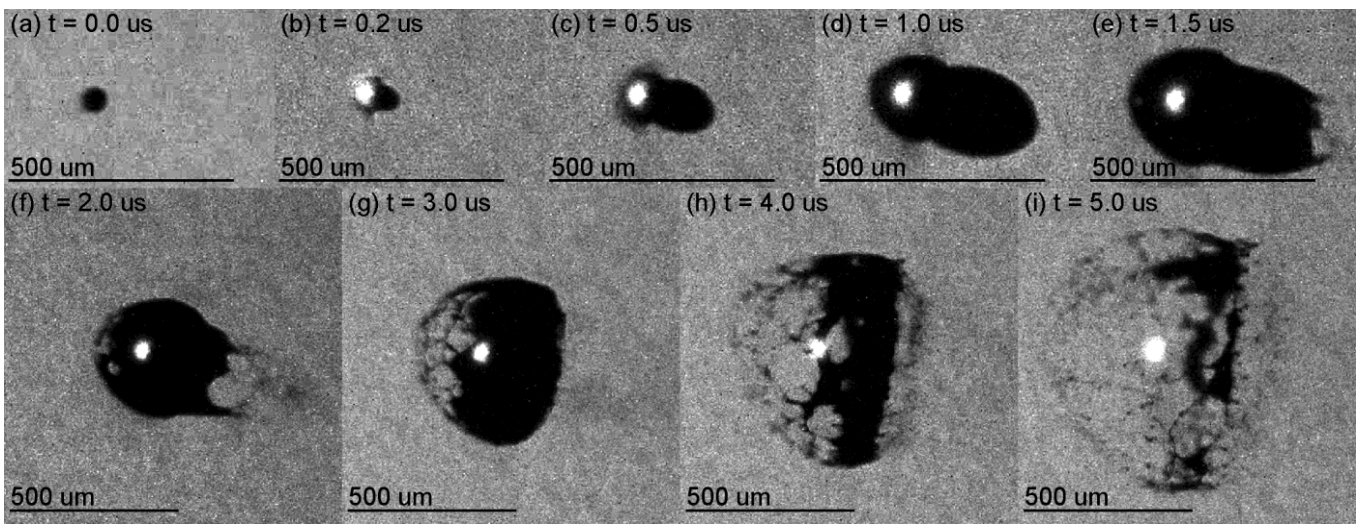

FIG. 1. The shadowgraphs for 60 um droplet taken at different time delays (t) after the 5.3 ps laser pulse.

**First we examine the evolution of the droplet shape upon the laser pulse impact for a fixed set of pulse parameters.** Note that further in text we refer to the deformed droplets as targets. Figure 1 shows a series of side-view shadowgraphs demonstrating the initial droplet and the resulting target shapes at the following moments of time after the impact: 0.2, 0.5, 1.0, 1.5, 2.0, 3.0, 4.0 and 5.0 us. The laser pulses propagate from left to right in the plane of figure at a slight (17˚) angle to horizon. The shadowgraphs are made at the following experimental conditions: initial droplet diameter $d_0$ = 60 um, E = 2.3 mJ, τ = 5.3 ps, focal spot diameter 72 um ($1/e^2$). The bright white spot which is present on all shadowgraphs corresponds to the light emission by the plasma close to the moment of laser pulse captured by the camera due to the long exposure. The series of shadowgraphs presented in Fig. 1 shows that the target undergoes a rapid expansion and deformation and finally fragments. The target deformation at the first stages is accompanied by the material ejection from the part of the target surface that was exposed to laser radiation (see Figs. 1(b)-(d)). The material ejection is represented on the shadowgraphs by the gray-to-black color gradient which can be interpreted as a dense vapor or mist of microparticles.

The target deformation proceeds in the following way. The initial spherical droplet is transformed during the first microsecond into a complex-shaped structure (see Figs. 1(a)-(d)) which can be approximated by two conjunct coaxial spheroids. Below in the text the part of the deformed target which expands around the initial droplet position is referred to as the forepart. Another part of the target, the elongated ellipsoidal part, is referred to as the rear part. During its further expansion the deformed target fragments. The fragmentation first occurs on the rear side between 1.0-1.5 us time, as depicted in Fig. 1(e). The rear part of the target completely fragments i.e. its shadow image disappears in the time range between 2.0 and 3.0 us (see Figs. 1(f)-(g)). The forepart of the deformed target fragments considerably slower. Its fragmentation starts from the front point of the forepart and propagates further towards the position of the constriction between the two parts of the target. As a result of the fragmentation, the residual target becomes ring-shaped (in our interpretation) as it is depicted in Fig. 1(i).

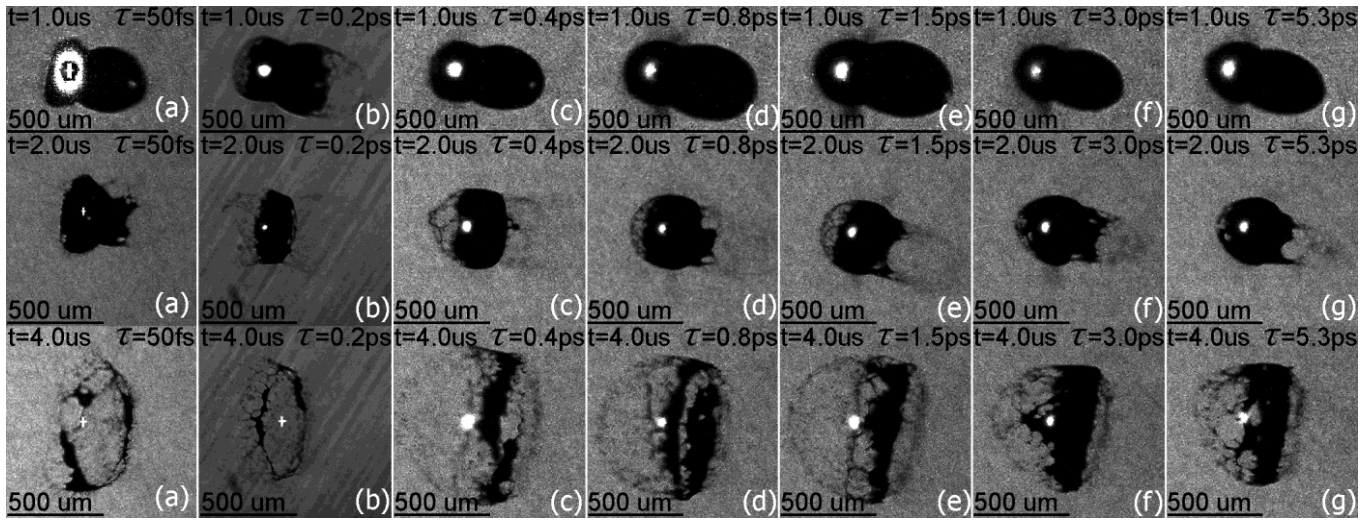

FIG. 2. Shadowgraphs for 60 um droplet taken at t = 1 us (upper row), t = 2 us (middle row), t = 4 us (lower row) for different pulse durations: (a) 50 fs, (b) 0.2 ps, (c) 0.4 ps, (d) 0.8 ps, (e) 1.5 ps, (f) 3.0 ps, (g) 5.3 ps.

**Now we consider the effect of laser pulse duration on the dynamics of the target shape evolution.** Figure 2 shows the side-view shadowgraphs demonstrating the target shape evolution after the laser impact for different pulse durations. Other experimental conditions are the same as on Fig. 1. As can be seen from Fig. 2(a) transformation of the target shape upon the impact proceeds in a similar way for all studied pulse durations. To be more precise, all targets are first transformed into structures composed of two expanding conjunct convex (ellipsoidal) parts (Fig. 2, upper row), then the targets are fragmented (Fig. 2, middle row) and finally the residual target materials takes a ring-shaped form (Fig. 2, lower row). However the dynamics of the target fragmentation significantly changes when switching to ultra-short pulses (see Figs. 2(a)-(b)). For pulse durations equal and above 0.2 ps the fragmentation of the rear part starts before the fragmentation of the forepart. Oppositely, for 0.05 ps pulse the fragmentation of the forepart starts first.

Figs. 1 and 2 demonstrate that during the process of the target fragmentation its boundaries remain sharp similar to the boundaries of the ruptured liquid surface. This brings us to the assumption that most of the target material remains liquid after the laser impact and that the expanding target has a hollow structure. Due to the presence of two conjunct ellipsoidal parts in the target shape (Fig. 2, upper row) one can assume that there are two cavities separately formed in the target during its expansion. The constriction between the two parts might indicate the existence of a membrane which separates these cavities.

**Let us discuss the quantative characteristics of the target expansion.** Figure 3 graphically illustrates the way we define the expansion velocities. In this figure, the grey disk designates the initial droplet, the solid black curve designates boundary (contour) of the expanding target. We quantify expansion of the target by calculating the displacement of its boundaries with respect to the initial target. For simplicity, we only quantify expansion along and perpendicular to the laser beam axis by calculating the velocities of the rare part ($V_r$) and the forepart ($V_f$) of the target and the velocity of the radial expansion ($V_n$), as shown in Fig. 3(a). Figure 3(b) shows an example of the calculated $V_r$, $V_f$ and $V_n$ values versus time delay after the laser pulse impact. The data corresponds to the 5.3 ps laser pulse that is described in the caption of Fig. 1. Note that it is impossible to determine $V_f$ values for delays shorter than 1 us because of plasma radiation captured by camera (see Fig. 1). It is impossible to determine $V_r$ values for delays longer than 1.5 us because of the fragmentation of the rear part. From Fig. 3(b) it is seen that the target expansion accelerates during first 0.5 us and then it starts decelerating. Assuming hollow structure of the target, we suggest that the observed acceleration of the expansion could be caused by the pressure of tin vapor which might fill the cavities of the target. However it is hardly feasible to experimentally verify the presence of the gaseous phase inside the target. Regarding the deceleration, we assume it to be caused by the surface tension forces. To comment on that in more details, let us consider the motion of an arbitrary small surface element of the target. We denote area of this element as *S* and its thickness *h*. Deceleration of this surface element can be calculated through the Laplace pressure ($p_L = 2\sigma/R$) as

$$a = -F/m = -2p_L S/(\rho S h) = -4\sigma/(\rho h R), \quad (1)$$

where $\rho$ is the density of liquid tin, $\sigma$ - surface tension, R - radius of surface curvature, factor 2 is due to presence of two surfaces. At the same time the value of deceleration can be determined from the experimental data from the slope of the curves in Fig. 3. And thus having the measured values of a and R, one can estimate from the Eq. (1) the shell thickness values in the different parts of the expanding target. For the particular case given in Fig. 3. curvature radiuses of various parts of target are: $R_r$ = 65 um, $R_f$ = 90 um, $R_n$ = 55 um and deceleration values: $a_f$ = -12.4x$10^6$ m/s$^2$, $a_n$ = -7.2x$10^6$ m/s$^2$, $a_r$ = -1.5x$10^8$ m/s2). These bring us to estimation for thicknesses: $h_r \approx$ 30 nm, $h_f \approx$ 260 nm, $h_n \approx$ 730 nm. For the estimation of h values we used tabular values of $\sigma$ = 0.53 N/m and $\rho$ = 7.3 g/cm$^3$ for indium-tin alloy from [16]. The estimated target thickness significantly differs over the target, which indicates that mass distribution is not uniform. Assuming that a rare part has a uniform thickness, one can estimate its mass is < 10% of the original droplet mass. It means that mass is mostly concentrated in the forepart of target. Furthermore, as depicted above the thickness of forepart is greater near the neck. This leads us to the assumption that the target material in the forepart is mostly concentrated in the ring-shaped region. Which is also confirmed via Figs. 1 and 2.

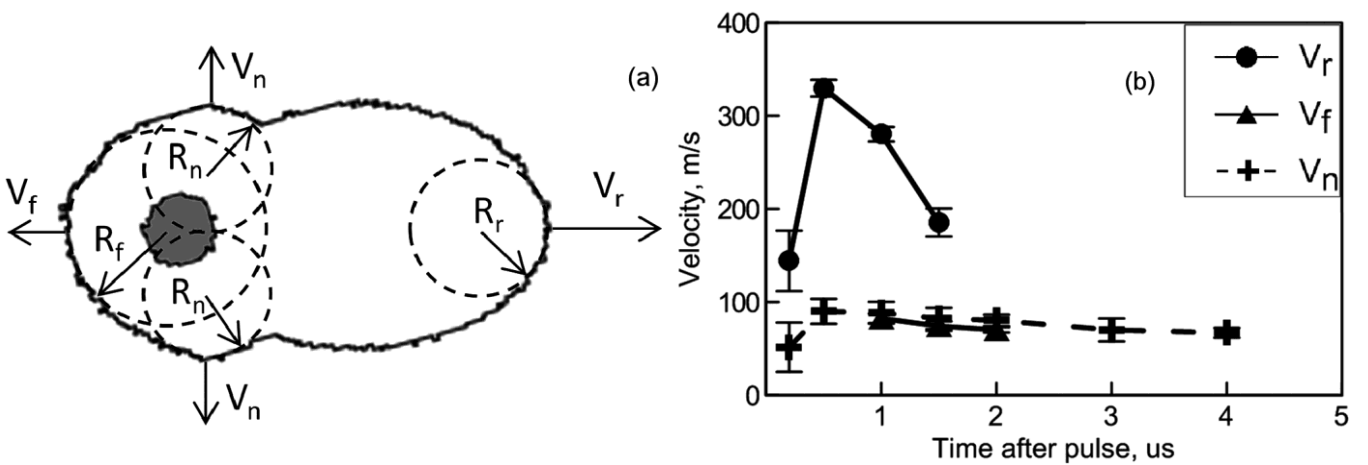

FIG. 3. (a) Boundaries of 60 um droplet at 0.0 and 1.0 us after 5.3 ps laser pulse. (b) Evolution of the target boundaries velocities.

**The scenario of the target shape evolution upon the laser impact reported above results to our understanding from the phenomena related to the propagation of shockwaves through liquids.** The average fluence on the

target surface in our experiments is about 100 J/cm$^2$ that significantly exceed the ablation threshold for tin. Depending on the pulse duration the irradiance varied in the range $10^{13} - 10^{15}$ W/cm$^2$. Laser energy is deposited in the skin depth surface layer of the droplet, which is a few tens nanometer thickness. Heat removal through the thermal conductivity to the bulk of the target is negligible at the picosecond and sub picosecond time scales. As a result all absorbed laser energy is spent on the ablation of the droplet material, plasma layer formation and its heating. According to Eidmann et al [15] the plasma pressure can reach values of Mbar scale at the considered here regimes of laser irradiances. Hence the laser impact results in the formation of the extremely strong shock at the surface of the irradiated hemisphere of the droplet.

The described above target shape evolution is a result of two phenomena. The first phenomenon is the cavitation in liquids caused by the convergent shockwaves [17]. The considered here laser-target configuration provides quasi-hemispherical geometry of the shockwave. And as result, the shockwave converges during its propagation through the body of the spherical target. When shockwave converge its magnitude increases and thus temperature and density of compressed matter also increases. This leads to a hot zone near the center of droplet which in turn leads to the cavitation and the formation of the target forepart shell. The second phenomenon is the so-called cavitating spall - the cavitation occurring when the shockwave reaches the opposite side of the liquid target and the rarefaction wave is reflected back [18]. This phenomenon is a cause of the formation of the rear part of the target. The combination of these two phenomena results in the observed target shapes which can be interpreted as two conjunct spheroid shells.

To conclude, we examined the hydrodynamic response of the liquid metal droplets to ultra-short and intense laser pulse impact. The observed evolution of the droplet shape upon the impact dramatically differs from the previously reported for nanosecond laser pulses, which result in flattening of the droplets. In our experiments the droplets undergo rapid asymmetric expansion and are then transformed into a complex shape which can be interpreted as two conjunct spheroid shells. The deformed liquid structures are eventually fragmented. We explain the described hydrodynamic response to the ultra-short impact as a result of the propagation of the laser-induced convergent shockwave through the volume of droplet.

**Acknowledgements**

We acknowledge M. M. Basko (Keldysh Institute of Applied Mathematics RAS), N. Inogamov (Landau Institute for Theoretical Physics RAS) and V. V. Zhakhovskii (Joint Institute for High Temperatures RAS) for the fruitful discussions.. The work was supported by the Ministry of Education and Science of the Russian Federation under the Agreement No 14.579.21.0004 (a unique identifier for Applied Scientific Research (project) RFMEFI57914X0004).